\documentclass{moriond}
\usepackage{multirow}
\usepackage{graphicx}
\usepackage{mathrsfs}
\usepackage{amsmath}
\def\Journal#1#2#3#4{{#1} {\bf #2}, #3 (#4)}

\def\be{\begin{equation}}
\def\ee{\end{equation}}
\def\bea{\begin{eqnarray}}
\def\eea{\end{eqnarray}}

\newcommand{\Photo}{\includegraphics[height=35mm]{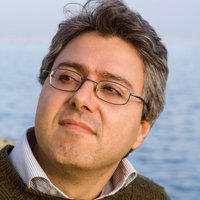}}
\begin{document}
\vspace*{4cm}
\title{Top quark production at ATLAS and CMS}

\author{Luca Lista,\\ \scriptsize{on behalf of the ATLAS and CMS collaborations}}

\address{INFN Sezione di Napoli, Comp. Univ. M. S. Angelo, via Cintia, 80126, Naples, Italy}

\maketitle

\abstracts{
A review of the main recent results on top quark production from the ATLAS and CMS experiments is presented. Results on both electroweak single top quark production and strong top pair production are presented.}

\section{Introduction}

Top quark production proceeds at hadron colliders via strong or electroweak processes.
The former results in the production of a top-antitop pair, while the latter 
results in the production of a single top quark or antiquark,
and may proceed via a W exchange in the $t$ or $s$ channel, or in associated production
with a W boson.
Top-pair production at LHC is enhanced compared to Tevatron thanks to the larger
gluon-fusion contribution. 
While at Tevatron the channel with the lowest cross section was the associated tW
production, which has been only observed at LHC, at LHC the single-top $s$ channel
has the smallest cross section, and is the hardest channel to probe.

\section{Single-top production}

\begin{figure}[htpb]
\centerline{\includegraphics[width=0.7\linewidth]{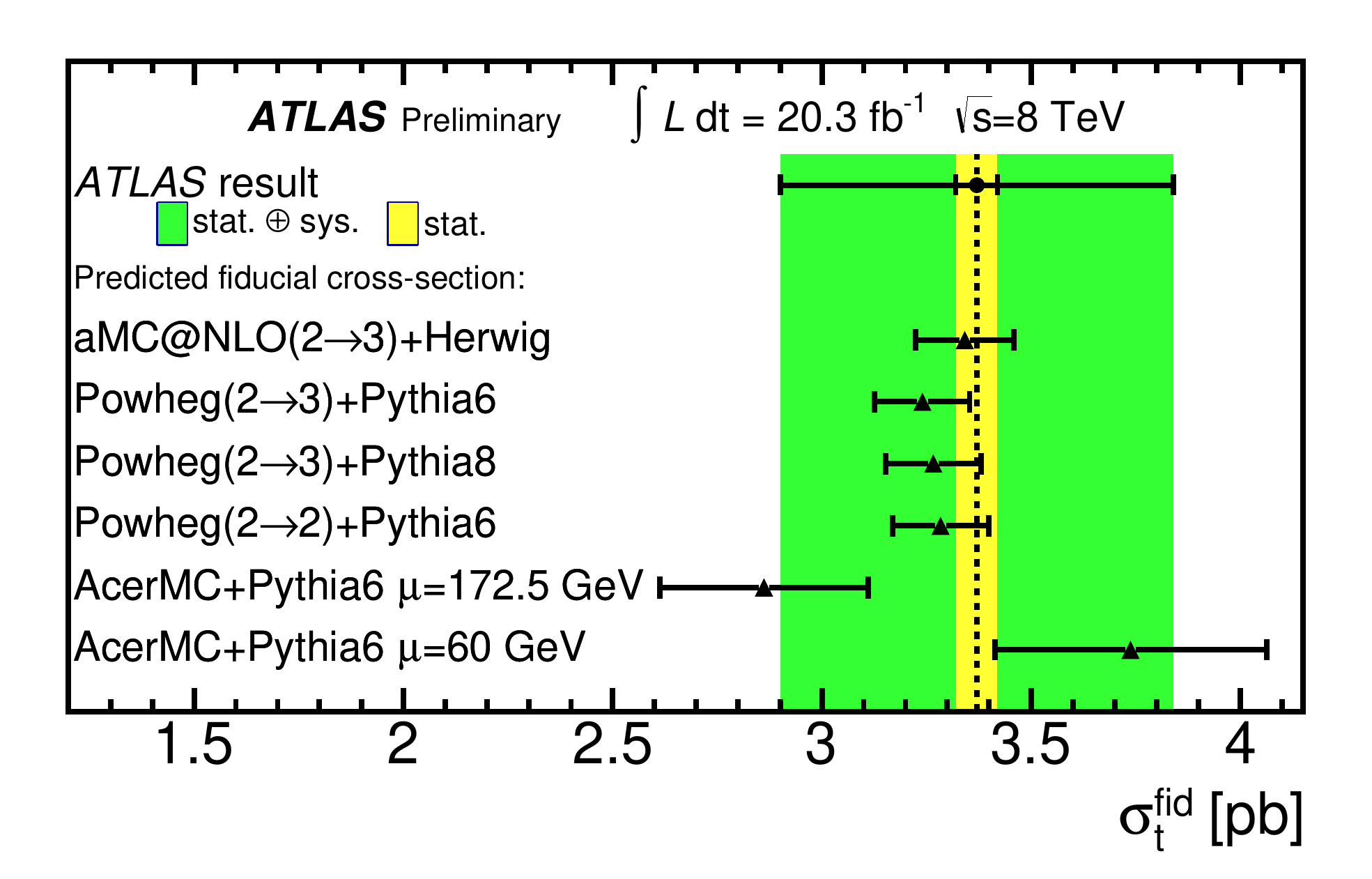}}
\caption[]{Comparison of ATLAS measurement of fiducial single-top production cross section
with different Monte Carlo generators.}
\label{fig:atlas_tch}
\end{figure}

\begin{figure}[htpb]
\centerline{\includegraphics[width=0.7\linewidth]{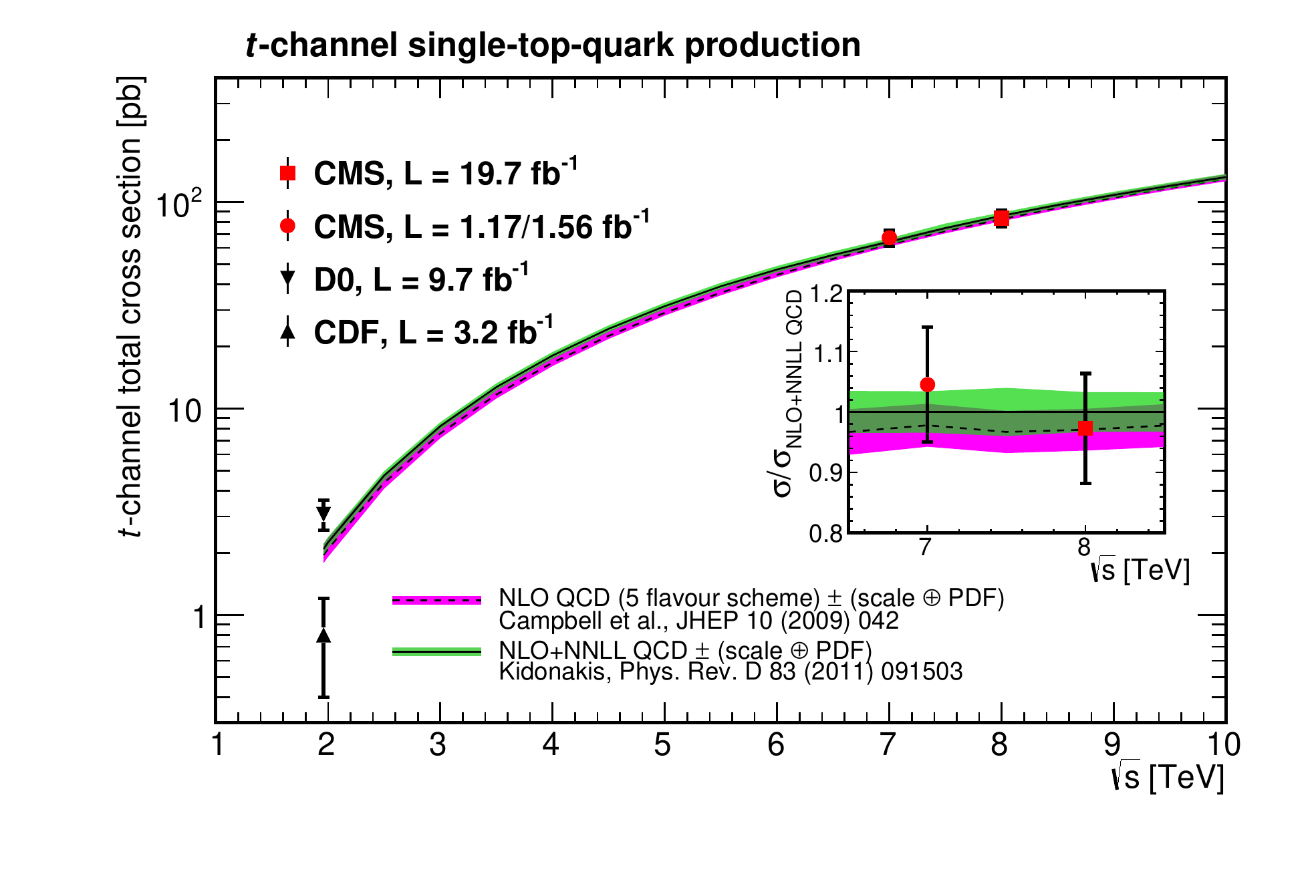}}
\caption[]{Comparison of CMS measurements of $t$-channel single-top production cross sections
at 7 and 8 TeV with next-to-leading order theory predictions.}
\label{fig:cms_tch}
\end{figure}

Single top quark production in the $t$ channel has been measured by ATLAS and CMS
both at 7 and 8 TeV. The most recent ATLAS analysis based on an integrated luminosity
of 20~fb$^{-1}$ collected at 8 TeV~\cite{atlas_t} adopted a neural network discriminant
to separate the $t$-channel signal from the backgrounds, using 14 discriminating 
variables. One electron or muon is required together with two hardonic jets, one of
which has been identified as stemming from a b quark. The multijet background is the 
hardest to model in simulation, and has been determined from data using a fit
to the distribution of the transverse missing energy.
ATLAS measures a fiducial cross section within the acceptance corresponding to a kinematic
selection specified in Ref.~\cite{atlas_t} to be:
\begin{equation}
\sigma_{t\text{-ch.}}^{\text{fid.}}= 3.37\pm 0.05\text{(stat.)}\pm 0.47\text{(syst.)}\pm 0.09\text{(lum.)}~\text{pb}\,,
\end{equation}
in agreement with theory predictions from different Monte Carlo (MC) generators, as
displayed in Fig.~\ref{fig:atlas_tch}.
The fiducial cross-section measurement can be extrapolated to the entire phase space
using acceptance estimates from generators. Assuming the prediction of the {\sc aMC@NLO} generator~\cite{amcnlo} plus
{\sc Herwig} parton shower~\cite{herwig}, the inclusive single-top 
production cross section is:
\begin{equation}
\sigma_{t\text{-ch.}}= 82.6\pm 1.2\text{(stat.)}\pm 11.4\text{(syst.)}\pm 
3.1\text{(pdf)}\pm 2.3\text{(lum.)}~\text{pb}\,,
\end{equation}
in agreement with next-to-leading order (NLO) theory predictions.
CMS measured the single-top production cross section in the $t$ channel
from a fit to the distribution of the pseudorapidity of the light jet accompanying
the top quark~\cite{cms_t}. 
The shapes of the distributions for the W+jets and the $\mathrm{t}\bar{\mathrm{t}}$
backgrounds are determined from control samples in data.
Events are selected requiring the presence of one electron or muon
together with tho jets, one of which is compatible with a b jet, and 
the reconstructed top-quark mass must be in the range $130-220$~GeV.
The inclusive production cross section determined by CMS is:
\begin{equation}
\sigma_{t\text{-ch.}}= 83.6\pm 2.3\text{(stat.)}\pm 7.4\text{(syst.)}~\text{pb}\,.
\end{equation}
The result is in agreement with NLO theory predictions,
as shown in Fig.~\ref{fig:cms_tch}, together with a previous CMS measurement at 7~TeV.
CMS also determined the ratio of cross sections at 8 and 7 TeV to be:
\begin{equation}
R_{8/7} = 1.24\pm 0.08\text{(stat.)}\pm 0.12\text{(syst.)}\,,
\end{equation}
and the ratio of top and antitop production cross section to be:
\begin{equation}
\sigma_{t\text{-ch.}\,, \mathrm{t}}/\sigma_{t\text{-ch.},\, \bar{\mathrm{t}}} = 
1.95\pm 0.10\text{(stat.)}\pm 0.19\text{(syst.)}\,.
\end{equation}
A measurement of the same ratio at 7~TeV by ATLAS gave~\cite{atlas_tat}:
\begin{equation}
\sigma_{t\text{-ch.}}(\mathrm{t})/\sigma_{t\text{-ch.}}(\bar{\mathrm{t}}) = 
1.81^{+0.23}_{-0.22}\,.
\end{equation}

The measurement of associated tW production is limited by the large 
 $\mathrm{t}\bar{\mathrm{t}}$ background. Both ATLAS and CMS use multivariate
techniques based on boosted decision trees (BDT) to extract the tW signal.
CMS analysis based on 12.2~fb$^{-1}$ collected at 8 TeV delivered the
first observation of tW~\cite{cms_tw} production with a cross section of:
\begin{equation}
\sigma_{\mathrm{tW}} = 24.5^{+5.5}_{-5.4}\,\text{pb}
\end{equation}
with a significance of 6.1 standard deviations (5.4 expected). ATLAS 
analysis based on 20.3~fb$^{-1}$ has a precision similar to CMS~\cite{atlas_tw}:
\begin{equation}
\sigma_{\mathrm{tW}} = 27.2\pm 2.8 \text{(stat.)}\pm 5.4\text{(syst.)}\,\text{pb}\,,
\end{equation}
and the quoted significance is 4.2$\sigma$ (4.0 expected).

At LHC the single-top channel with the lowest production cross section is the 
$s$ channel. CMS searched for the $s$-channel production in the entire sample 
at 8 TeV. No significant excess over the expected background was observed, and an upper
limit on the production cross sectoin in the $s$ channel was set at the
95\% confidence level (CL):
\begin{equation}
\sigma_{s\text{-ch.}} < 11.5~\text{pb}\,.
\end{equation}
This limit corresponds to 2.1 times the standard model (SM) cross section.
The largest uncertainty on the above limit is due to the theory modelling
of the large  $\mathrm{t}\bar{\mathrm{t}}$ background. The use of
more recent NLO MC generators to simulate this background is expected
to improve the precision of future measurements of this very rare channel.
ATLAS published the limit on $s$-channel production at 7 TeV based on 
the first 0.7~fb$^{-1}$:
\begin{equation}
\sigma_{s\text{-ch.}} < 26.5~\text{pb}\,,
\end{equation}
corresponding to 5.8 times the SM cross section value.

\subsection{$|V_{\mathrm{tb}}|$ determination}

Single-top production allows a direct probe of the
tWb copling at the production vertex. In particular, assuming that
the branching fraction of the top quark to Wb is equal to one, the
square root of the measured single-top production cross section,
divided by its theory prediction, computed assuming $|V_{\mathrm{tb}}|=1$, 
must be equal to $|V_{\mathrm{tb}}|$ in the SM. Deviations from such value may be
indications of new physics that modifies the tWb vertex coupling.
The available measurements of single-top production in the $t$ and
tW channels allow determinations of $|V_{\mathrm{tb}}|$ with different levels of
precision. CMS also quoted a combination of the $|V_{\mathrm{tb}}|$
measurements using the two $t$-channel cross section measurements at
7 and 8 TeV, which achieves the best present precision (4.1\%).
Table~\ref{tab:vtb} reports the $|V_{\mathrm{tb}}|$
determinations from the available measurements.

\begin{table}[t]
\caption[]{$|V_{\mathrm{tb}}|$ measurements from single-top production cross section.}
\label{tab:vtb}
\vspace{0.4cm}
\begin{center}
\begin{tabular}{|l|c|c|c|c|}
\hline
Experiment & $\sqrt{s}$ & Process & $|V_{\mathrm{tb}}|$ & Precision\\
\hline\hline
 \multirow{4}{*}{ATLAS} &  \multirow{2}{*}{7 TeV} & $t$-ch. & $1.13^{+0.14}_{-0.13}$ & 11.9\%\\ \cline{3-4}
 & & tW & $1.03^{+0.16}_{-0.19}$ & 17.0\%\\ \cline{2-4}
 & \multirow{2}{*}{8 TeV} & $t$-ch. & $0.97^{+0.09}_{-0.10}$ & 9.8\%\\ \cline{3-4}
 & & tW & $1.10\pm 0.12\text{(exp.)}\pm 0.03\text{(th.)}$ & 11.2\%\\ 
\hline
 \multirow{5}{*}{CMS}&  \multirow{2}{*}{7 TeV} & $t$-ch. & $1.020\pm 0.046\text{(exp.)}\pm 0.017\text{(th.)}$ & 4.8\%\\ \cline{3-4}
 & & tW & $1.01^{+0.16}_{-0.13}\text{(exp.)}^{+0.03}_{-0.04}\text{(th.)}$ & 14.8\% \\ \cline{2-4}
 & \multirow{2}{*}{8 TeV} & $t$-ch. & $0.979\pm 0.045\text{(exp.)}\pm 0.016\text{(th.)}$ & 4.9\%\\ \cline{3-4}
 & & tW & $1.03\pm 0.12\text{(exp.)}\pm 0.04\text{(th.)}$ & 12.3\%\\ \cline{3-4}
 & 7+8 TeV &  $t$-ch. &  $0.998\pm 0.038\text{(exp.)}\pm 0.016\text{(th.)}$ & 4.1\%\\
\hline
\end{tabular}
\end{center}
\end{table}

\subsection{New physics with single top}
Single-top production at LHC offers the opportunity to look for new physics that could
affect the tgu, tgc, tZu or tZc vertices, exhibiting anomalous couplings, compared
to the SM prediction. Such anomalous couplings would result in flavour-changing neutral currents (FCNC).
ATLAS studied FCNC at the top production vertex~\cite{atlas_fcnc}, which would result in a 
single top quark produced without any associated particles. The study based on the 8~TeV data sample set limits
to anomalous tgu and tgc coupling that can be translated into limits to the FCNC branching fractions of the top quark:
\begin{eqnarray}
{\cal B}(\mathrm{t}\rightarrow\mathrm{gu}) & < & 3.1\times 10^{-5}\,, \\
{\cal B}(\mathrm{t}\rightarrow\mathrm{gc}) & < & 1.6\times 10^{-5}\,,
\end{eqnarray}
at the 95\% CL. CMS looked for FCNC associated production of a top quark and a Z boson~\cite{cms_fcnc}, which resulted in 
limits to anomalous tZu and tZc compling, translated into the following branching ratio limits
at the 95\% CL:
\begin{eqnarray}
{\cal B}(\mathrm{t}\rightarrow\mathrm{Zu}) & < & 5.1\times 10^{-3}\,, \\
{\cal B}(\mathrm{t}\rightarrow\mathrm{Zc}) & < & 0.114\,.
\end{eqnarray}

\section{Top-antitop pair production}

\begin{figure}[htpb]
\centerline{\includegraphics[width=0.6\linewidth]{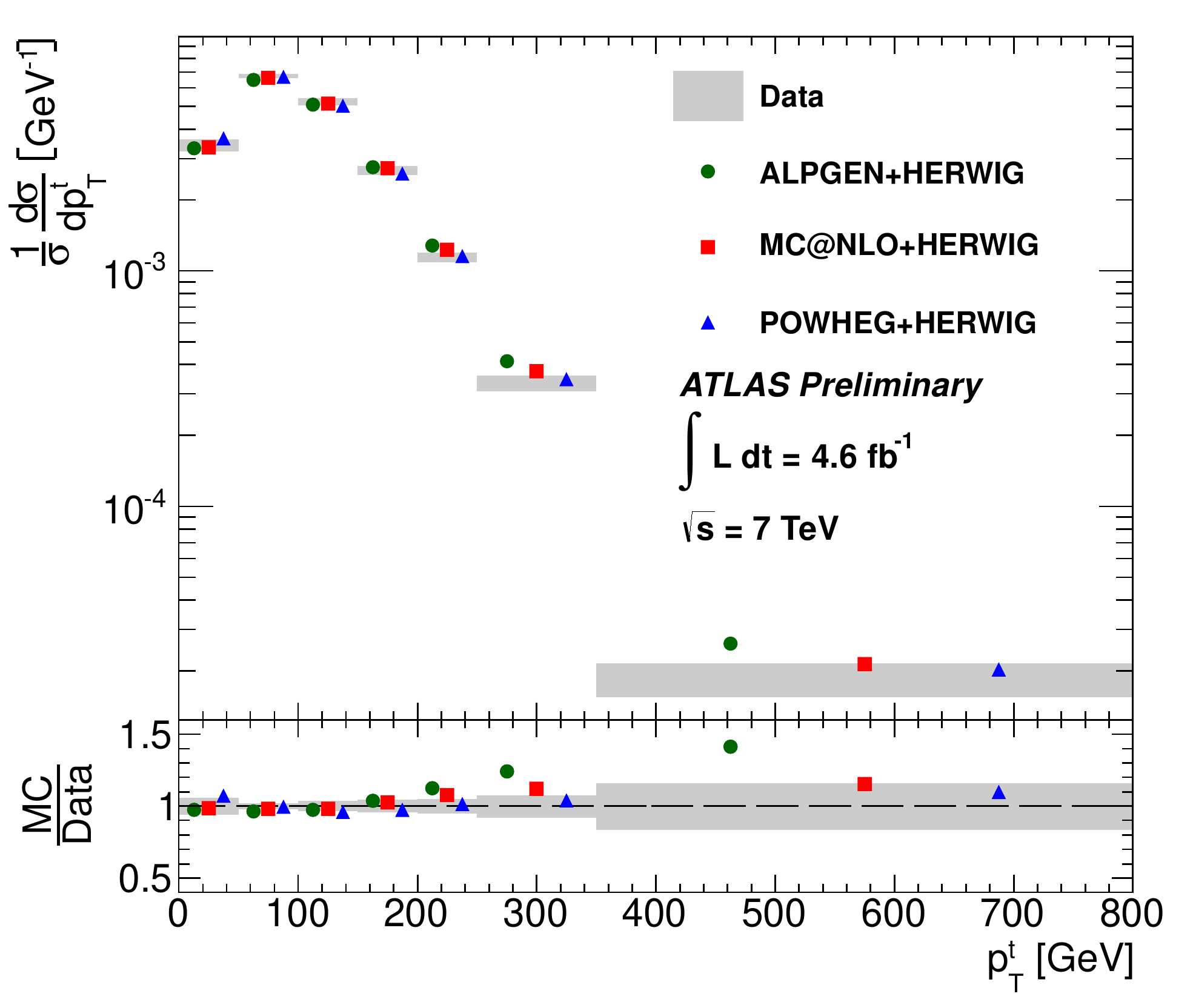}}
\caption[]{Differential $\mathrm{t}\bar{\mathrm{t}}$ cross section measurement by ATLAS
as a function of the top-quark transverse momentum, compared with prediction
of different Monte Carlo generators.}
\label{fig:atlas_diff_pt}
\end{figure}

\begin{figure}[htpb]
\centerline{\includegraphics[width=0.6\linewidth]{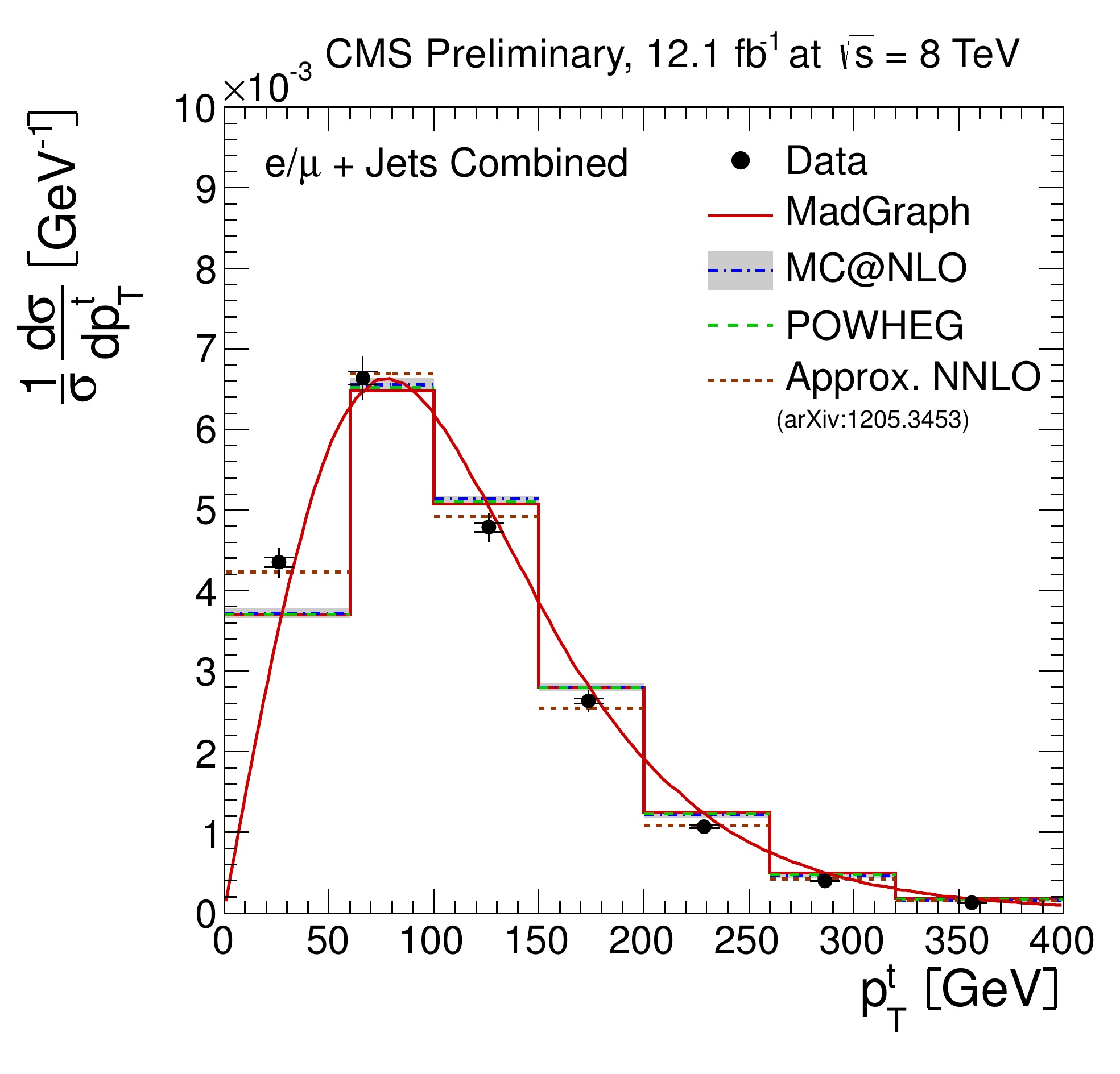}}
\caption[]{Differential $\mathrm{t}\bar{\mathrm{t}}$ cross section measurement by CMS
as a function of the top-quark transverse momentum, compared with prediction
of different Monte Carlo generators and approximate next-to-next-to-leading-order prediction.}
\label{fig:cms_diff_pt}
\end{figure}

The most precise determinations of $\mathrm{t}\bar{\mathrm{t}}$ production at LHC
is performed using dileptonic events. CMS updated the analysis at 8~TeV using
a data sample of 5.3~fb$^{-1}$, obtaining the following cross section measurement~\cite{cms_tt}:
\begin{equation}
\sigma_{\mathrm{t}\bar{\mathrm{t}}} = 239\pm 2\text{(stat.)}\pm 11\text{(syst.)}\pm 6\text{(lum.)}\,\text{pb}
\end{equation}
ATLAS analyzed the entire 8~TeV dataset implementing an in-situ determination of the 
b-tagging efficiency from the b-tag multiplicity distribution, and achieved the following
result~\cite{atlas_tt}:
\begin{equation}
\sigma_{\mathrm{t}\bar{\mathrm{t}}} = 237.7\pm 1.7\text{(stat.)}\pm 7.4\text{(syst.)}\pm 7.4\text{(lum.)}\,.
\pm 4.0\text{(beam energy)}\,\text{pb}
\end{equation}
Inclusive cross-section measurements at LHC are in good agreement with the SM prediction,
and the large $\mathrm{t}\bar{\mathrm{t}}$ data sample allows studies of differential
distribution of several observable quantities, in order to perform more stringent
SM tests and to validate detailed features of MC generators. The observed variables
include the transverse momentum and rapidity of top quarks and $\mathrm{t}\bar{\mathrm{t}}$ pair,
the invariant mass of the $\mathrm{t}\bar{\mathrm{t}}$ pair, additional jet multiplicity, and more.
Experimental effects are unfolded from the raw observed spectra.
Most of the observed distribution show good ageement with generator prediction.
Among the few cases where discrepancies have been observed, Fig.~\ref{fig:atlas_diff_pt}
shows that not all MC generators correccly describe the distribution of the 
top-quark transverse momentum at large $p_{\mathrm{T}}$ values, as reported by
ATLAS~\cite{atlas_diff} at 7 TeV.
CMS reports at 8 TeV~\cite{cms_diff} a deficit in the leading-order and  next-to-leading-order MC predictions
at low top-quark $p_{\mathrm{T}}$, while the approximate next-to-NLO predictions correctly describe
the data. 
This is shown in Fig.~\ref{fig:cms_diff_pt}.

\section{$\mathrm{t}\bar{\mathrm{t}}$ associated production}

The production of $\mathrm{t}\bar{\mathrm{t}}$ pairs associated with more particles,
either jets or vector bosons, can be studied at LHC exploiting the large integrated luminosity accumulated
during the run at 7 and 8 TeV. CMS measured the production of $\mathrm{t}\bar{\mathrm{t}}$
plus a $\mathrm{b}\bar{\mathrm{b}}$ pair~\cite{cms_ttbb}, which is an important background
to ttH and new physics searches. Applying two different thresholds to the transverse momenta
of the reconstructed jets, CMS measured the ratio of $\mathrm{t}\bar{\mathrm{t}}\mathrm{b}\bar{\mathrm{b}}$
to $\mathrm{t}\bar{\mathrm{t}}$ plus two jets at 8~TeV:
\begin{eqnarray}
\sigma_{\mathrm{t}\bar{\mathrm{t}}\mathrm{b}\bar{\mathrm{b}}}/\sigma_{\mathrm{t}\bar{\mathrm{t}}jj} & =
0.023\pm 0.003\text{(stat.)}\pm 0.005\text{(syst.)}\,, & p_{\mathrm{T}} > 20~\text{GeV}\,, \\
\sigma_{\mathrm{t}\bar{\mathrm{t}}\mathrm{b}\bar{\mathrm{b}}}/\sigma_{\mathrm{t}\bar{\mathrm{t}}jj} & = 
0.022\pm 0.004\text{(stat.)}\pm 0.005\text{(syst.)}\,, & p_{\mathrm{T}} > 40~\text{GeV}\,.
\end{eqnarray}
Both measurements are compatible with the prediction of {\sc MadGraph}~\cite{madgraph} and {\sc Powheg}~\cite{powheg} 
generators, though both have central values slightly larger than the SM predictions.
ATLAS measured at 7~TeV the incusive production cross section of 
 $\mathrm{t}\bar{\mathrm{t}}$ plus a b or c quark~\cite{atlas_ttbc}:
\begin{equation}
\sigma_{\mathrm{t}\bar{\mathrm{t}}+\mathrm{b}/\mathrm{c}+X}/
\sigma_{\mathrm{t}\bar{\mathrm{t}}+\ge 1 jet} =
6.2\pm 1.1\text{(stat.)}\pm 1.8\text{(syst.)}\,,
\end{equation}
assuming the jet has $p_{\mathrm{T}}> 25$~GeV and $|\eta|<2.5$.
This result is also in agreement with the SM predictions, using {\sc AlpGen}~\cite{alpgen},
though, as for the previous CMS measurement, the central value is somewhat larger than the 
theory prediction.

\begin{figure}[htpb]
\centerline{\includegraphics[width=0.6\linewidth]{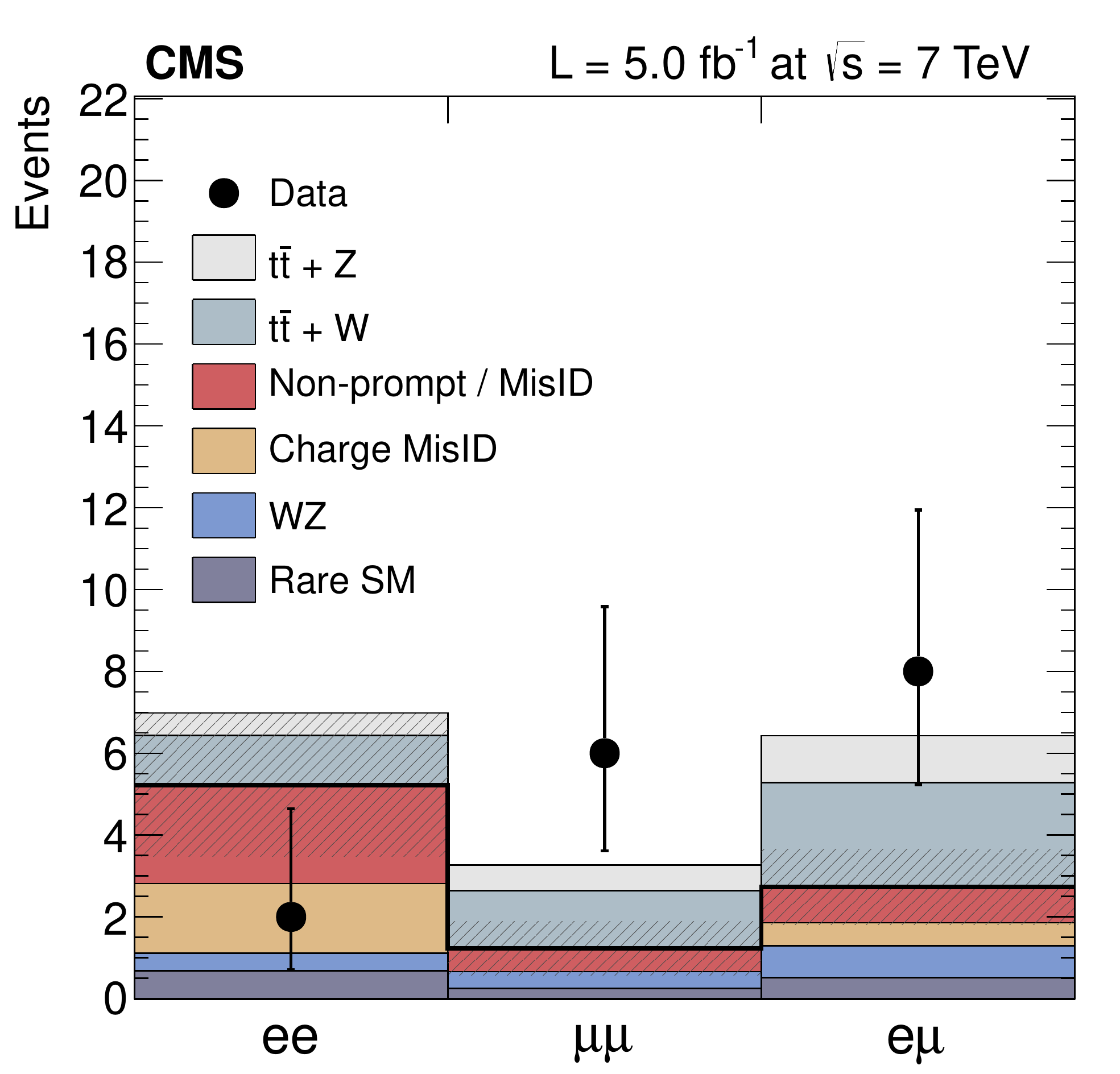}}
\caption[]{Expected and measured event yields in the CMS same-sign dilepton analysis for  $\mathrm{t}\bar{\mathrm{t}}W$, $\mathrm{t}\bar{\mathrm{t}}Z$ and the various backgrounds, superimposed.}
\label{fig:cms_ttv}
\end{figure}

Measurements of $\mathrm{t}\bar{\mathrm{t}}$ associated with W or Z were performed by CMS
at 7~TeV~\cite{cms_ttv} using two analyses: an inclusive search of 
$\mathrm{t}\bar{\mathrm{t}}$ plus either W or Z in events with same-sign lepton pairs (Fig.~\ref{fig:cms_ttv}),
and an exclusive analysis of events with three leptons, looking for $\mathrm{t}\bar{\mathrm{t}}$Z.
The results have a limited precision, due to the limited number of selected events:
\begin{eqnarray}
\sigma_{\mathrm{t}\bar{\mathrm{t}}V} & = & 0.43^{+0.17}_{-0.15}\text{(stat.)}^{+0.09}_{-0.07}\text{(syst.)}\,\text{pb}\,,\\
\sigma_{\mathrm{t}\bar{\mathrm{t}}\mathrm{Z}} & = & 0.28^{+0.14}_{-0.11}\text{(stat.)}^{+0.06}_{-0.03}\text{(syst.)}\,\text{pb}\,.
\end{eqnarray}
The two measurements have statistical significance of 3.0 and 3.3 standard deviations respectively, and
are in agreement with the SM predictions.

\begin{figure}[htpb]
\centerline{\includegraphics[width=0.8\linewidth]{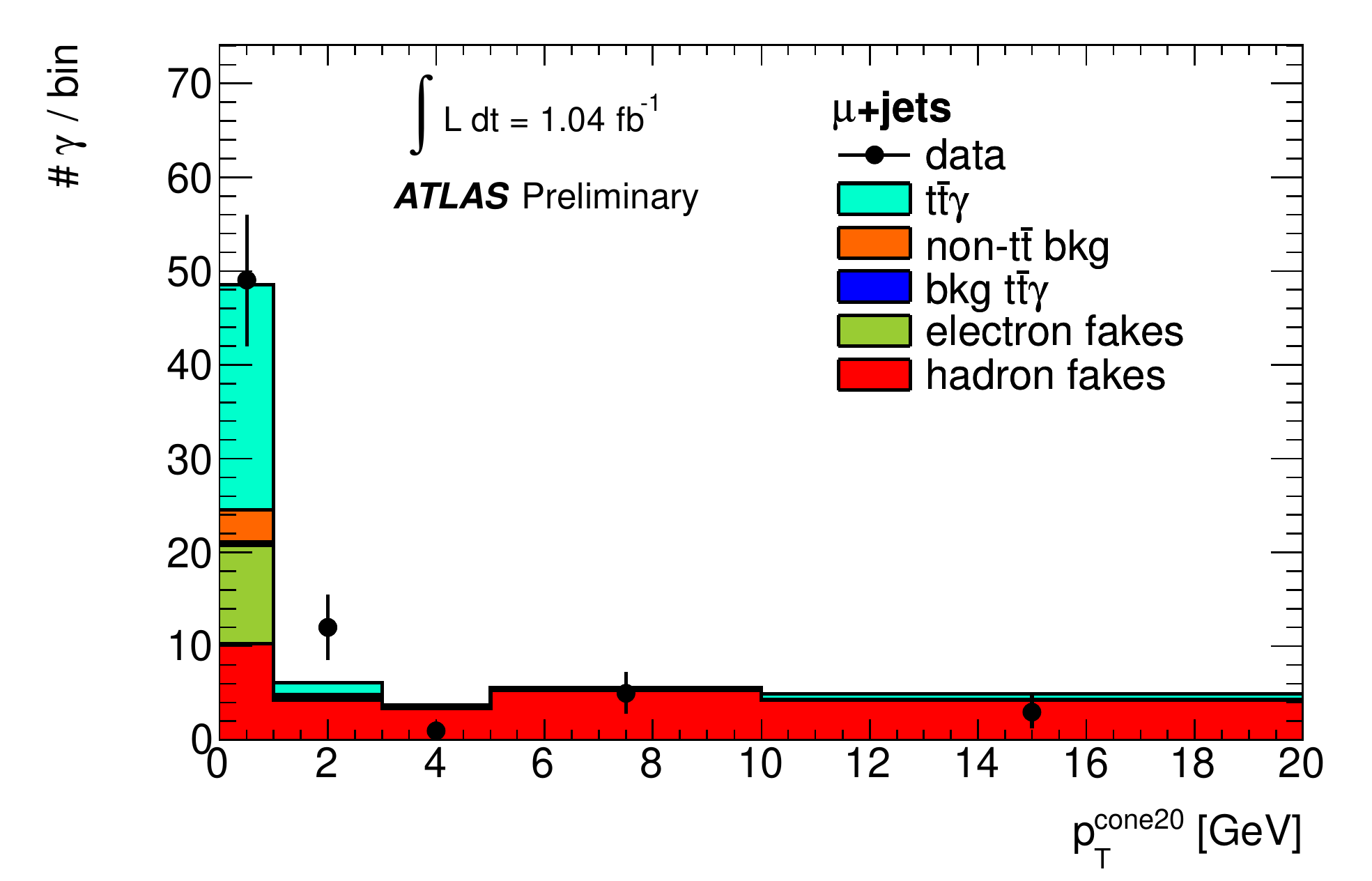}}
\caption[]{Result of the isolation fit for the single muon channel. The predicted $tt\gamma$ signal is shown on top of the different background contributions. }
\label{fig:atlas_ttg}
\end{figure}

ATLAS measured $\mathrm{t}\bar{\mathrm{t}}\gamma$ production at 7~TeV in events with a lepton and a jet~\cite{atlas_ttg}.
Photon fake rate is difficult to model in simulation, and has been determined from data using a fit to the 
distribution of a variable describing the photon's isolation (Fig.~\ref{fig:atlas_ttg}).
The result is:
\begin{equation}
\sigma_{\mathrm{t}\bar{\mathrm{t}}\gamma} = 2.0\pm 0.5\text{(stat.)}\pm 0.6\text{(syst.)}\pm 0.1\text{(lum.)}\,\text{pb}\,,
\end{equation}
assuming a photon transverse energy greater than 8~GeV.
CMS updated the $\mathrm{t}\bar{\mathrm{t}}\gamma$ measurement at 8 TeV, and the result is:
\begin{equation}
\sigma_{\mathrm{t}\bar{\mathrm{t}}\gamma}/\sigma_{\mathrm{t}\bar{\mathrm{t}}} = 
(1.07\pm 0.07\text{(stat.)}\pm 0.27\text{(syst.)})\times 10^{-2}\,,
\end{equation}
assuming a photon transverse energy greater than 20~GeV, and a separation between the photon
and the b jet from the top-quark decay $\Delta R>0.1$, which can be translated into:
\begin{equation}
\sigma_{\mathrm{t}\bar{\mathrm{t}}\gamma} = 2.4\pm 0.2\text{(stat.)}\pm 0.6\text{(syst.)}\,\text{pb}\,.
\end{equation}

CMS searched for the production of two top-antitop quark pairs at 8 TeV. This process has a
cross section of the order of one fb in the SM~\cite{tttt_th1}$^,\,$\cite{tttt_th2}, 
but could be enhanced in several models beyond the SM. The CMS analysis looks for the decay
of one top quark with an electron or muon, and the hadronic decay of the other three top-quark.
In order to determine the three-jet combinations that have large probability to come form a
top quark decay, dedicated BDTs are trained. A second BDT is used to select signal events adding
event variables. The analysis showed no excess over the expected background, and the following
upper limit at 95\% CL has been set~\cite{cms_tttt}:
\begin{equation}
\sigma_{\mathrm{t}\bar{\mathrm{t}}\mathrm{t}\bar{\mathrm{t}}} < 63~\text{fb}\,.
\end{equation}

\section{Conclusions}
The precision of top production measurements at LHC is steadily improving, and the
focus of ATLAS and CMS is shifting towards the precise understanding of top production mechanism
in order to perform detailed comparisons with state-of-the-art QCD predictions, and
possibly find deviations from the standard model. The determination of cross sections
in fiducial regions is being adopted in order to avoid model-dependent extrapolations
that would introduce theory uncertainty.
Next updates of LHC activities in top physics will involve both targeting the ultimate precision
for upcoming 7 and 8 TeV run-I legacy measurements, and the preparation for
run-II data at higher LHC energy.


\section*{References}

\begin{thebibliography}{99}

\bibitem{atlas_t} ATLAS Collaboration, ATLAS-CONF-2014-007.
\bibitem{amcnlo} J. Alwall {\it et al.}, \Journal{JHEP}{128}{1106}{2011}.
\bibitem{herwig} G. Corcella {\it et al.}, \Journal{JHEP}{010}{0101}{2001}.
\bibitem{cms_t} CMS Collaboration, arXiv:1403.7366, sub. to JHEP.
\bibitem{atlas_tat} ATLAS Collaboration, ATLAS-CONF-2012-056.
\bibitem{cms_tw} CMS Collaboration, arxiv:1401.2942 , sub. to Phys. Rev. Lett.
\bibitem{atlas_tw} ATLAS Collaboration, ATLAS-CONF-2013-100.
\bibitem{atlas_fcnc} ATLAS Collaboration, ATLAS-CONF-2013-063.
\bibitem{cms_fcnc} CMS Collaboration, CMS PAS TOP-12-021.
\bibitem{cms_tt} CMS Collaboration, \Journal{JHEP}{02}{024}{2014}.
\bibitem{atlas_tt} ATLAS Collaboration, ATLAS-CONF-2013-097.
\bibitem{atlas_diff} ATLAS Collaboration, ATLAS-CONF-2013-099.
\bibitem{cms_diff} CMS Collaboration, CMS-PAS TOP-12-027.
\bibitem{cms_ttbb} CMS Collaboration, CMS-PAS-TOP-13-010.
\bibitem{madgraph} F. Maltoni and T. Stelzer, \Journal{JHEP}{02}{027}{2003}.
\bibitem{powheg} S. Alioli {\it et al.}, \Journal{JHEP}{06}{043}{2010}.
\bibitem{atlas_ttbc} ATLAS Collaboration, \Journal{Phys. Rev.}{D89}{072012}{2014}.
\bibitem{alpgen} M.L. Mangano {\it et al.,} \Journal{JHEP}{7}{001}{2003}.
\bibitem{cms_ttv} CMS Collaboration, \Journal{Phys. Rev. Lett.}{110}{1720}{2013}.
\bibitem{atlas_ttg} ATLAS Collaboration, ATLAS-CONF-2011-153.
\bibitem{tttt_th1} V. Barger {\it et al.}, \Journal{Phys. Lett.}{B687}{70}{2010}.
\bibitem{tttt_th2} M.W.G. Bevilacqua, \Journal{JHEP}{1207}{111}{2012}.
\bibitem{cms_tttt} CMS Collaboration, CMS PAS TOP-13-012.
\end{thebibliography}
\end{document}